\documentclass[12pt,preprint]{aastex}
\usepackage{emulateapj5}
\newcommand{\mincir}{\raise
-2.truept\hbox{\rlap{\hbox{$\sim$}}\raise5.truept\hbox{$<$}\ }}
\newcommand{\magcir}{\raise
-2.truept\hbox{\rlap{\hbox{$\sim$}}\raise5.truept\hbox{$>$}\ }}
\newcommand{\minmag}{\raise
-2.truept\hbox{\rlap{\hbox{$<$}}\raise6.truept\hbox{$<$}\ }}
 
\usepackage{graphicx}
\newcommand{\be}{\begin{equation}}
\newcommand{\ee}{\end{equation}}

\newenvironment{inlinefigure}{%
\def\@captype{inlinefigure}%
\noindent\begin{minipage}{\linewidth}\begin{center}}
{\end{center}\end{minipage}\smallskip}
\makeatother
\shorttitle{Breaking the $\sigma_{8}-\Omega_m$ degeneracy}
\shortauthors{S. Basilakos \& M. Plionis}

\begin{document}


\title{Breaking the $\sigma_{8}-\Omega_m$ degeneracy using the
  clustering of high-$z$ X-ray AGN}
\author{Spyros Basilakos\altaffilmark{1} and Manolis Plionis\altaffilmark{2,3}}
\altaffiltext{1}{Research Center for Astronomy \& Applied Mathematics,
Academy of Athens, Soranou Efessiou 4, GR-11527 Athens, Greece}
\altaffiltext{2}{Institute of Astronomy \& Astrophysics, 
National Observatory of Athens, Palaia Penteli 152 36, Athens, Greece}
\altaffiltext{3}{Instituto Nacional de Astrof\'{\i}sica \'Optica y
Electr\'onica, AP 51 y 216, 72000, Puebla, Pue, M\'exico}

\begin{abstract}
The clustering of X-ray selected AGN appears to be a valuable tool
for extracting cosmological information. 
Using the 
recent high-precision angular clustering results of $\sim 30000$ XMM-{\it
Newton} soft (0.5-2\,keV) X-ray sources (Ebrero et al.), which
have a median redshift of $z\sim 1$, and assuming a flat geometry, 
a constant in comoving coordinates AGN clustering evolution and the AGN bias
evolution model of Basilakos et al.,  
we manage to break the $\Omega_m-\sigma_8$ degeneracy.
The resulting cosmological constraints are: 
$\Omega_m=0.27^{+0.03}_{-0.05}$, 
w$=-0.90^{+0.10}_{-0.16}$ and 
$\sigma_{8}=0.74^{+0.14}_{-0.12}$, while the dark matter host halo
mass, in which the X-ray selected AGN are presumed to reside, is 
$M=2.50^{+0.50}_{-1.50}\times 10^{13}h^{-1}M_{\odot}$. 
For the constant $\Lambda$ model (w=$-1$) we find
$\Omega_m=0.24 \pm 0.06$
and $\sigma_{8}=0.83^{+0.11}_{-0.16}$, in good agreement with
recent studies based on cluster abundances, weak lensing
 and the CMB, but in disagreement with the recent bulk flow
analysis.

\end{abstract}

\vspace{1.0cm}

\section{Introduction}

A large variety of cosmologically relevant data, based on the
combination of galaxy clustering, the supernova Ia's
Hubble relation, the cosmic microwave background
(CMB) fluctuations and weak-lensing strongly support a flat universe,
containing cold dark matter (CDM) and ``dark energy'' which
is necessary to explain 
the observed accelerated cosmic expansion
(eg., 
Komatsu et al. 2010; Hicken et al. 2009; Fu et al. 2008 and references therein). 

The nature of the mechanism that is responsible for the late-time 
acceleration of the Hubble expansion is a fundamental problem in modern
theoretical physics and cosmology. Due to the absence of a physically 
well-motivated fundamental theory, various proposals
have been suggested in the literature, among which 
a cosmological constant, a time varying vacuum 
quintessence, $k-$essence, vector fields,
phantom, tachyons, Chaplygin gas, etc
(eg., 
Weinberg 1989; 
Peebles \& Ratra 2003; 
Boehmer \& Harko 2007; Padmanabhan 2008 and references therein). 
Note that the simplest pattern of dark energy corresponds to 
a scalar field having a self-interaction potential with the associated
field energy density decreasing with a slower rate than the
matter energy density. In such case the
dark energy component is described by an equation of state
$p_{Q}={\rm w}\rho_{Q}$ with w$<-1/3$
(dubbed ``quintessence'', eg. Peebles \& Ratra 2003 and references
therein). The traditional cosmological constant 
($\Lambda$) model corresponds to w$=-1$.
The viability of the different dark-energy models in reproducing the current excellent
cosmological data and the requirements of galaxy formation is a
subject of intense work (eg. Basilakos, Plionis \& Sol\'a 2009 and
references therein).

Another important cosmological parameter is the normalization
of the cold dark matter power spectrum in the form of
the rms density fluctuations in spheres of radius 8$h^{-1}$ Mpc, the
so-called $\sigma_8$. There is a degenerate relation between $\sigma_8$ and
$\Omega_m$ (eg. Eke, Cole \& Frenk 1996; Wang \& Steinhardt 1998; 
Henry et al. 2009; Rozo et al. 2009 and references therein) and it 
is important to improve current constraints in order to break such degeneracies.
Furthermore, there are also apparent inconsistencies between the values 
of $\sigma_8$ provided by different observational methods, 
among which the most deviant and problematic for the {\em concordance} cosmology,
is provided by the recent bulk flow analysis 
of Watkins, Feldman \& Hudson (2009).

In this paper we extend our previous work (Basilakos \& Plionis 2009;
hereafter BP09), using the angular clustering of the largest sample of
high-$z$ X-ray selected active galactic nuclei (Ebrero et al. 2009a), 
in an attempt to break the $\sigma_{8}-\Omega_m$ degeneracy within
spatially flat cosmological models.

\section{Basic Methodology} 
The main ingredients of the method used to put
cosmological constraints based on the angular clustering of some extragalactic
mass-tracer, has been already presented in our previous papers (see
also Matsubara 2004; BP09 and references
therein). It consists in
comparing the observed angular clustering with that predicted by
different primordial fluctuations power-spectra, using
Limber's equation to invert from spatial to angular clustering. By minimizing the
differences of the observed and predicted angular correlation function,
one can constrain the cosmological parameters that enter
in the power-spectrum determination as well as in Limber's inversion.
Below we 
present only the main steps of the procedure.

\subsection{Theoretical Angular and Spatial Clustering} 
Using the well known Limber's inversion equation (Limber 1953), we can relate the
angular and spatial clustering of any extragalactic population under
the assumption of power-law correlations and the small angle approximation
(see details in BP09). After some algebraic calculations 
and within the context of flat spatial geometry, we can
easily write the angular correlation function as:
\begin{equation}
\label{eq:angu}
w(\theta)=2\frac{H_{0}}{c} \int_{0}^{\infty} 
\left(\frac{1}{N}\frac{{\rm d}N}{{\rm d}z} \right)^{2}E(z){\rm d}z 
\int_{0}^{\infty} \xi(r,z) {\rm d}u \;\;,
\end{equation} 
where ${\rm d}N/{\rm d}z$ is the source redshift distribution,
estimated by integrating the appropriate source luminosity function
(in our case that of Ebrero et al. 2009b), folding in also the area
curve of the survey. We also have
\be 
E(z)=[\Omega_{m}(1+z)^{3}+(1-\Omega_{m})(1+z)^{3(1+{\rm
    w})}]^{1/2}\;,
\ee
with w the dark-energy equation of state parameter given by
$p_{Q}={\rm w}\rho_{Q}$ with w$<-1/3$.
The source spatial correlation function is: 
\be
\xi(r,z) = (1+z)^{-(3+\epsilon)}b^{2}(z)\xi_{\rm DM}(r)\;,
\ee 
where $b(z)$ is the evolution of the linear bias
factor, $\epsilon$ is a parameter related to the model
of AGN clustering evolution (eg. de Zotti et al. 1990)\footnote{
Following
K\'undic (1997) and Basilakos \& Plionis (2005; 2006) 
we use the
constant in comoving coordinates clustering model, ie., $\epsilon=-1.2$.}
and $\xi_{\rm DM}(r)$ 
is the corresponding correlation function of the underlying dark
matter distribution, given by the Fourier transform of the 
spatial power spectrum $P(k)$
of the matter fluctuations, linearly
extrapolated to the present epoch: 
\be
\label{eq:spat1}
\xi_{\rm DM}(r)=\frac{1}{2\pi^{2}}
\int_{0}^{\infty} k^{2}P(k) 
\frac{{\rm sin}(kr)}{kr}{\rm d}k \;\;.
\ee
We use the nominal functional form of 
the CDM power spectrum, $P(k)=P_{0} k^{n}T^{2}(k)$, with
$T(k)$ the CDM transfer function 
(Bardeen et al. 1986; Sugiyama 1995)
and $n\simeq 0.96$, following the 5 (and 7)-year WMAP results (Komatsu et
al. 2010), and a baryonic
density of $\Omega_{\rm b} h^{2}= 0.022 (\pm 0.002)$. The
normalization of the power-spectrum, $P_{0}$, can be parametrized by
the rms mass fluctuations on $R_{8}=8 h^{-1}$Mpc scales ($\sigma_8$), 
according to:
\be
P_{0}=2\pi^{2} \sigma_{8}^{2} \left[ \int_{0}^{\infty} T^{2}(k)
 k^{n+2} W^{2}(kR_{8}){\rm d}k \right]^{-1} \;,
\ee
where 
$W(kR_{8})=3({\rm sin}kR_{8}-kR_{8}{\rm cos}kR_{8})/(kR_{8})^{3}$. 
Regarding the Hubble constant we use either 
$H_{0}\simeq 71$ kms$^{-1}$Mpc$^{-1}$ 
(Freedman 2001; Komatsu et al. 2010)
or $H_{0}\simeq 74$ kms$^{-1}$Mpc$^{-1}$ (Riess et al. 2009).
Note, that in the current analysis 
we also utilize the non-linear corrections 
introduced by Peacock \& Dodds (1994).  

\subsection{X-ray AGN bias evolution}
The notion of the bias between mass-tracers and underlying DM mass is an
essential ingredient for CDM models in order to reproduce the observed
extragalactic source 
distribution (eg. Kaiser 1984; Davis et al. 1985; Bardeen et
al. 1986)
Although a large number of models have been proposed in
 the literature to model the evolution of the bias factor, 
in the current analysis we use our own approach, 
which was described initially
in Basilakos \& Plionis (2001; 2003) and extended in Basilakos,
Plionis \& Ragone-Figueroa (2008; hereafter BPR08). 

For a benefit of the reader we remind that our bias model 
is based on linear perturbation theory and the 
Friedmann-Lemaitre solutions of the cosmological
field equations, while it also 
allows for interactions and merging of the mass tracers.
Under the usual assumption that each X-ray AGN is 
hosted by a dark matter halo of the same mass, we can present analytically
its bias evolution behavior. 
A more realistic view, however, of the AGN host halo having a 
spread of masses around a given value, with a 
given distribution that does not change 
significantly with redshift, should not alter 
the predictions of our bias evolution model.

For the case of a spatially flat
cosmological model, our bias evolution model predicts:
\be
b(M,z)={\cal C}_{1}(M)E(z)+{\cal C}_{2}(M)E(z)I(z)+y(z)+1 \;,
\label{bias1}
\ee
where

\be
y(z)= E(z) \left[ \int_{0}^{z}\frac{{\cal K}(x)I(x) {\rm d}x}{(1+x)^{3}}-
I(z)\int_{0}^{z} \frac{{\cal K}(x) {\rm d}x}{(1+x)^{3}} \right]
\ee
with ${\cal K}(z)=f(z) E^{2}(z)$, 
$I(z)=\int_{z}^{\infty} (1+x)^{3}  {\rm d}x /E^{3}(x)$,
\be\label{eq:ff}
f(z)=A(m-2) (1+z)^{m} E(z)/D(z)\;,
\ee
\be\label{eq:fitC}
{\cal C}_{1,2}(M)\simeq \alpha_{1,2} (M/10^{13}h^{-1}M_{\odot})^{\beta_{1,2}} \;,
\ee
The various constants are given in BPR08\footnote{For the benefit of the 
reader we present the values of the constants for a few selected cases:
$\alpha_{1}=3.29$, $\beta_{1}=0.34$
and $\alpha_{2}=-0.36$, $\beta_{2}=0.32$, while in eq.(\ref{eq:ff}) we have
$A=5\times 10^{-3}$ and $m=2.62$ for $1\times 10^{13} \le M/h^{-1}M_{\odot} \le 3\times
10^{13}$, and $A=6\times10^{-3}$ and $m=2.54$ for the case of 
$1\times 10^{13}< M/h^{-1}M_{\odot} \le 6\times 10^{13}$.}.
Note that $D(z)$ is the linear growth factor (scaled to unity at the
present time), useful expressions of which can be found for the 
dark energy models in Silveira \& Waga (1994) and in Basilakos (2003). 

\begin{inlinefigure}
\plotone{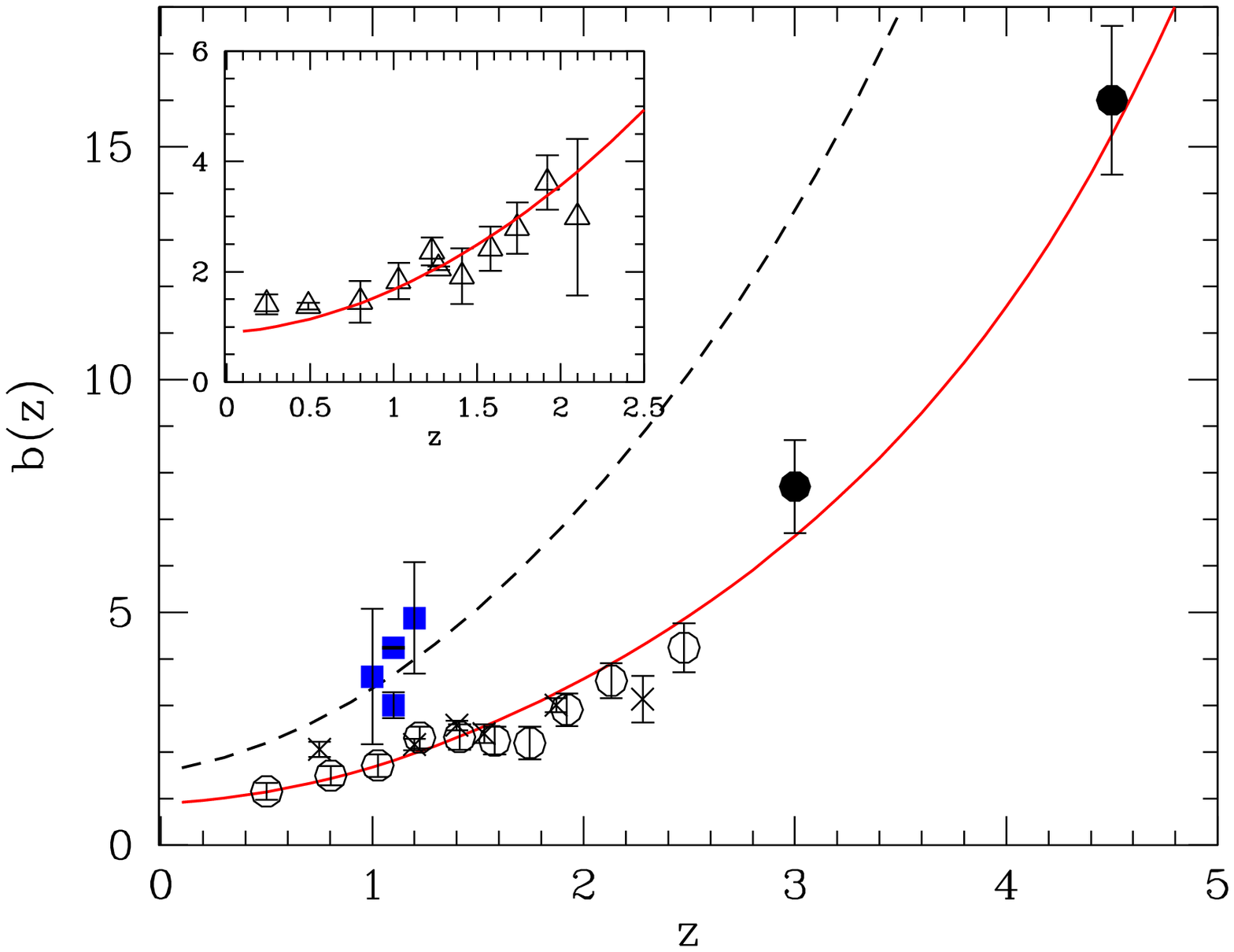}
\figcaption{The observed evolution of AGN bias (different points) compared 
with the BPR08 model predictions (curves). 
Optically selected SDSS and 2dF quasars are represented by empty dots
and crosses, respectively, while X-ray selected AGN by filled (blue) squares.
In the insert we plot the most recent optical QSO bias values based on the
SDSS quasar uniform sample (Ross et al. 2009). 
The curves represent the expectations of the BPR08 model, with the solid
(red) lines corresponding to a DM halo mass of
$M=10^{13} \;h^{-1} \; M_{\odot}$ and the dashed line to
$M=2.5 \times 10^{13} \;h^{-1} \; M_{\odot}$.}
\end{inlinefigure}

In order to provide an insight on the success or failure of our bias 
evolution
model, we compare in Fig.1 the measured bias values of optical and
X-ray selected AGNs with our $b(z)$ model. 
The bias of optical quasars by Croom et al. (2005), Myers et al. (2007), 
Shen et al. (2007) and Ross et al. (2009)
based on the 2dF QSOs (open circles), 
SDSS DR4 (crosses), 
SDSS DR5 (solid points) and 
the SDSS quasar uniform sample (insert panel), 
are well approximated by
our $b(z)$ model for a DM halo of 
$10^{13}h^{-1}M_{\odot}$ (solid red line) in agreement with previous
studies (Porciani, Magliocchetti, Norberg 2004;
Croom et al. 2005; Negrello, Magliocchetti \& de Zotti 2006; 
Hopkins et al. 2007). However, what is worth stressing is that our
model is (to our knowledge) the only one that can simultaneously fit
the lower redshift ($z<2.5$) optical AGN bias with the higher ($z>3$) results of Shen
et al. (2007) for the same halo mass of $10^{-13}h^{-1}M_{\odot}$.
The solid (blue) squares represent the bias of the soft 
X-ray selected AGNs, based on a
variety of X-ray surveys (eg. 
Basilakos et al. 2005; Puccetti et al. 2006; 
Gilli et al. 2009; Ebrero et al. 2009a). 
The model $b(z)$ curve (dashed line) that fits 
these results correspond to halo
masses $M=2.5 \times 10^{13} \; h^{-1} \; M_{\odot}$, strongly
indicating that X-ray and optically selected AGN do not inhabit the
same DM halos.

\section{Cosmological Parameter Estimation}

We use the most recent measurement of the angular correlation
function of X-ray selected AGN (Ebrero et al. 2009a).
This measurement is based on a sample (hereafter 2XMM) constructed from
1063 XMM-{\it Newton} observations at high galactic latitudes 
and includes $\sim 30000$ soft (0.5-2\,keV) point sources within an effective
area of $\sim 125.5$ deg$^{2}$ and 
an effective flux-limit of $f_x \ge 1.4 \times
10^{-15}$ erg cm$^{-2}$ s$^{-1}$  (for more details 
see Mateos et al. 2008). 
Notice that the redshift selection function of the X-ray sources, 
obtained by using the soft-band luminosity function of Ebrero et al.
(2009b), that takes into account the realistic luminosity dependent density 
evolution of the X-rays sources, predicts a characteristic 
depth of $z\sim 1$. 

In BP09, using the 2XMM clustering, we already provided
stringent cosmological
constraints in the $\Omega_m-{\rm w}$ plane, using as a prior a flat
cosmology and the WMAP7 power-spectrum normalization value of Komatsu et
al. (2010). 
In the current analysis we relax the latter prior and
allow $\sigma_8$ to be a free parameter to
be fitted by the data. Therefore the 
corresponding free-parameter vector that enters
the standard $\chi^{2}$ likelihood procedure, which compares the observed 
and predicted clustering, is: ${\bf p} \equiv (\Omega_{m}, {\rm w}, \sigma_{8},M)$, 
with $M$ the AGN host dark matter halo mass, which enters in our BPR08 biasing
evolution scheme.

The likelihood estimator\footnote{Likelihoods
  are normalized to their maximum values.}, is defined as:
${\cal L}_{\rm AGN}({\bf p})\propto {\rm exp}[-\chi^{2}_{\rm AGN}({\bf p})/2]$
with:
\be
\label{eq:likel}
\chi^{2}_{\rm AGN}({\bf p})=\sum_{i=1}^{n} \left[ w_{\rm th}
(\theta_{i},{\bf p})-w_{\rm obs}(\theta_{i}) \right]^{2}/(
\sigma^{2}_{i}+\sigma^{2}_{\theta_{i}})  \;\;,
\ee 
where $n$ and $\sigma_{i}$ is the number of logarithmic bins ($n=13$) and 
the uncertainty of the observed angular correlation 
function respectively, while
$\sigma_{\theta_{i}}$ corresponds to the width
of the angular separation bins.

We sample the various parameters in a grid as follows:
the matter density $\Omega_{m} \in [0.01,1]$ in steps of
0.01; the equation of state parameter w$\in [-1.6,-0.34]$ in steps
of 0.01; the rms matter fluctuations $\sigma_{8} \in [0.4,1.4]$ in steps 
of 0.01 and the parent dark matter halo
$M/10^{13}h^{-1}M_{\odot} \in [0.1,4]$ in steps of 0.1.
Note that we have allowed the parameter w to take values below -1.

Our main results are listed in Table 1, where we quote the best 
fit parameters with the corresponding 1$\sigma$ uncertainties, 
for two different values of the Hubble
constant. Small variations around $\sim
71$ kms$^{-1}$Mpc$^{-1}$ (which is the value used in the rest of the paper), 
appear to provide statistically indistinguishable results.
The likelihood function of the soft X-ray sources peaks at
$\Omega_{m}=0.27^{+0.03}_{-0.05}$, w$=-0.90^{+0.11}_{-0.19}$, 
$\sigma_{8}=0.74^{+0.14}_{-0.12}$ and 
$M=2.5^{+0.5}_{-1.5}\times 10^{13}\;h^{-1}M_{\odot}$, with a reduced
$\chi^{2}$ of $\sim 4$. Such a large $\chi^2/$df value is caused by the
measured small
$w(\theta)$ uncertainties in combination with the observed $w(\theta)$
sinusoidal modulation (see BP09). Had we used a 2$\sigma$ $w(\theta)$
uncertainty in eq.(\ref{eq:likel}) 
we would have obtained roughly the same constraints and
a reduced $\chi^{2}$ of $\sim 1$. The apparent sinusoidal $w(\theta)$ modulation
is a subject of further investigation.

\begin{inlinefigure}
\epsscale{1.1}
\plotone{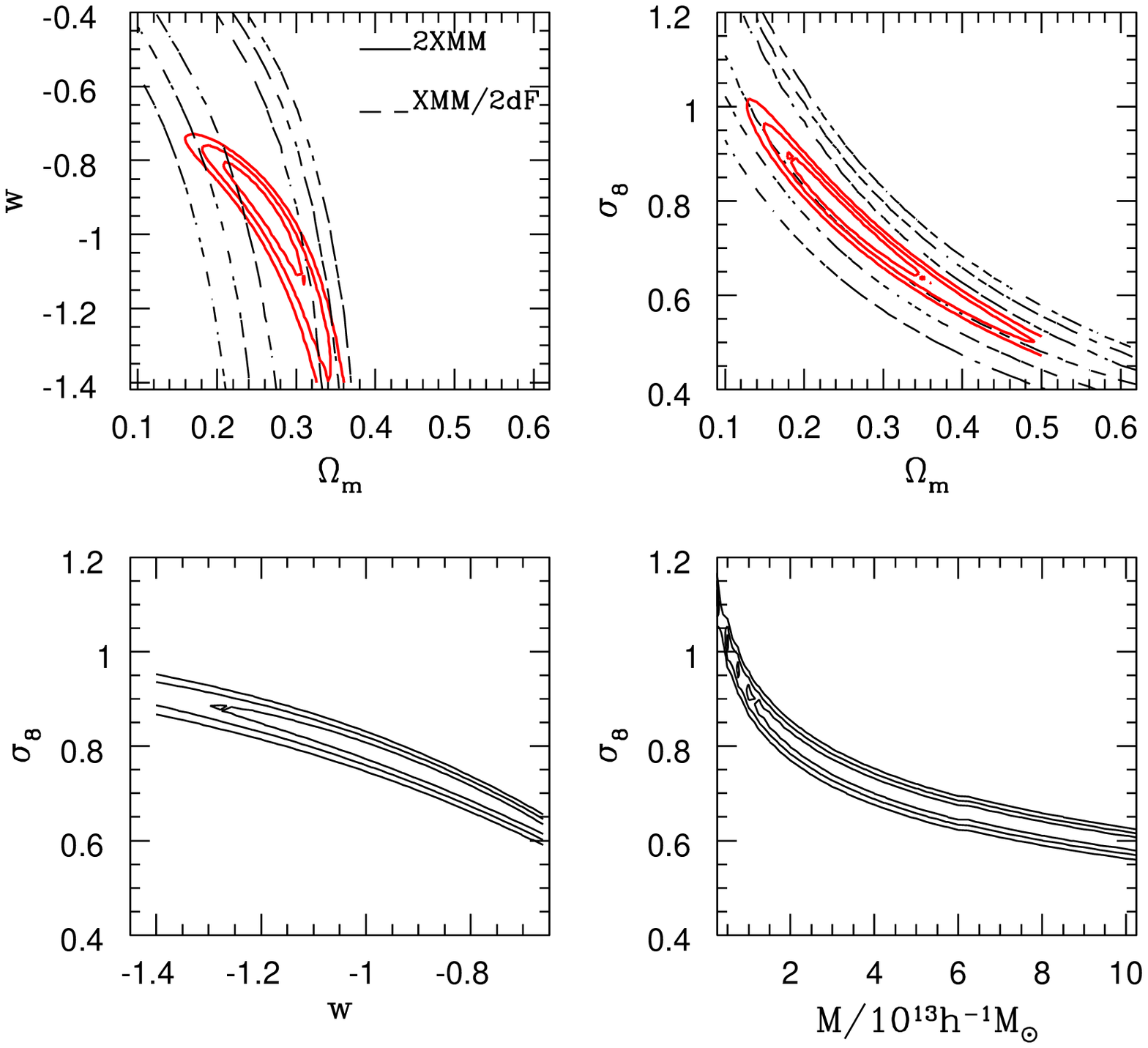}
\figcaption{Likelihood contours (1$\sigma$,
2$\sigma$ and $3\sigma$) in the following planes:
$(\Omega_m,{\rm w})$ (upper left panel), 
$(\Omega_m,\sigma_{8})$ (upper right panel),
$(\sigma_{8},{\rm w})$ (bottom left panel) 
and $(\sigma_{8},M)$ (bottom right panel). 
In the upper two panels we show for clarity our current solution 
with thick (red) contours, while the dashed contours correspond to our previous
analysis, based on the shallower XMM/2dF survey (Basilakos \& Plionis 2006). 
}
\end{inlinefigure}

In Fig.2 we present the 1$\sigma$, 2$\sigma$ and $3\sigma$
confidence levels (corresponding to where
$-2{\rm ln}{\cal L}/{\cal L}_{\rm max}$ equals 2.30, 6.16 and 11.83)
in the $(\Omega_m,{\rm w})$,
$(\Omega_m,\sigma_{8})$,
$(\sigma_{8},{\rm w})$ and $(\sigma_{8},M)$, 
planes, by marginalizing the first one over 
$M$ and $\sigma_{8}$, the second
one over $M$ and ${\rm w}$, the third one over
$M$ and $\Omega_m$ and the last
over $\Omega_m$ and ${\rm w}$.
We also present, with dashed lines, our previous solution
of Basilakos \& Plionis (2006), which where derived 
by using the 
shallower (effective flux-limit of 
$f_x \ge 2.7 \times10^{-14}$ erg cm$^{-2}$ s$^{-1}$)
and significantly smaller ($\sim 2.3$ deg$^{2}$) XMM/2dF survey 
(Basilakos et al. 2005). 
Comparing our current results with our previous analysis 
it becomes evident that with the current high-precision X-ray AGN correlation
function of Ebrero et al. (2009a) we have achieved to break the
$\Omega_m-\sigma_8$ degeneracy and to substantially improve the
constraints on $\Omega_m$, ${\rm w}$ and $\sigma_8$. However, 
there are still degeneracies, the most important of which is 
in the w$-\sigma_8$ plane.

It should be mentioned that some recent works, based on the large-scale bulk flows,
strongly challenge the {\em concordance} $\Lambda$CDM cosmology
by implying a very large $\sigma_8$ value. Indeed,
Watkins et al. (2009), using a variety of tracers to measure 
the bulk flow on scales of $\sim 100 \; h^{-1}$Mpc, found a value of $\sim 400$ km s$^{-1}$
that implies a $\sigma_{8}$ normalization which is a factor of $\sim 2$ larger than what expected 
in the {\em concordance} cosmology.
On the high $\sigma_8$ side are also the results of Reichardt et al. (2009), based on the 
secondary Sunayev-Zeldovich anisotropies in the CMB, providing $\sigma_{8}\simeq 0.94$, as well as a
novel analysis based on the integrated Sachs-Wolfe effect 
(Ho et al. 2008).

Contrary to the above results, our X-ray AGN clustering analysis provides a $\sigma_8$ value consistent
with the {\em concordance} cosmology and in agreement with a variety of other studies.
In particular, for w$=-1$ ($\Lambda$ cosmology) and
$M=2.5^{+0.3}_{-0.2}\times 10^{13}\;h^{-1}M_{\odot}$, we find
$\Omega_m=0.24\pm 0.06$ and $\sigma_{8}=0.83^{+0.11}_{-0.16}$
(see Table 1).
Our results are in agreement with those of recent cluster abundances studies, providing 
(for w$=-1$):
$\sigma_8= 0.86 \pm 0.04 (\Omega_m/0.32)^{-0.3}$
(Henry et al. 2009) and $\sigma_8= 0.83 \pm 0.03
(\Omega_m/0.25)^{-0.41}$ (Rozo et al. 2009).
Furthermore, Mantz et al. (2009) using as a new cosmological tool  
the simultaneous fit of the cosmological parameter space and the
cluster X-ray luminosity-mass relation, also broke the
$\Omega_m-\sigma_8$ degeneracy and found:
$\Omega_m=0.23\pm 0.04$, $\sigma_{8}=0.82\pm 0.05$ and
w$=-1.01\pm 0.20$ (see their Table 2 and Fig. 4), which are in 
excellent agreement with our w$=-1$ results (see last row in our Table 1).
Note, that their combined analysis (utilizing also
the CMB, baryonic acoustic oscillations and gas mass fraction) provides
$\Omega_m=0.27 \pm 0.02$, w$=-0.96 \pm 0.06$ and 
$\sigma_{8}=0.79 \pm 0.03$. 
Moreover, Fu et al. (2008) based on a weak-lensing analysis found 
the degenerate combination
$\sigma_{8}=0.837\pm0.084(\Omega_m/0.25)^{0.-53}$. 
From the peculiar velocities
statistical analysis, Pike \& Hudson (2005) and Cabr\'e \& Gazta\~naga
(2009) obtained $\sigma_{8}=0.88\pm 0.05(\Omega_m/0.25)^{-0.53}$ and 
$\sigma_{8}=0.85\pm 0.06$ (for $\Omega_m=0.245$), respectively. 
The consistency of all the (seven) previously mentioned works (including the
current study) can be also appreciated from their average $\sigma_8$ value which
is (for w$=-1$ and ignoring the different $\Omega_m$ values): 
$\langle \sigma_8 \rangle = 0.844 \pm 0.009$, 
where the quoted uncertainty is the 1$\sigma$ scatter of the mean. 
Note that the combined WMAP 7-years+SNIa+BAO analysis 
of Komatsu et al. (2010) 
provide  a slightly lower value of $\sigma_{8}=0.809 \pm
0.024$ (with $\Omega_m=0.272 \pm 0.015$ and w$=-0.98 \pm 0.05$).

\section{Conclusions}
We have used the recent angular clustering measurements
of high-$z$ X-ray selected AGN, identified as soft (0.5-2 keV) XMM point
sources (Ebrero et al. 2009a),
in order to break the degeneracy between 
the rms mass fluctuations $\sigma_{8}$ and $\Omega_m$. 
Applying a standard likelihood procedure, assuming a constant in
comoving coordinates AGN clustering evolution, the bias
evolution model of Basilakos et al. (2008) and 
a spatially flat geometry, we put relatively stringent constraints 
on the main cosmological parameters, given by: 
$\Omega_m=0.27^{+0.03}_{-0.05}$, w$=-0.90^{+0.10}_{-0.16}$ and 
$\sigma_{8}=0.74^{+0.14}_{-0.12}$. We also 
find that the dark matter host halo mass, in which the X-ray selected
AGN are assumed to reside, is 
$M=2.50^{+0.50}_{-1.50}\times 10^{13}h^{-1}M_{\odot}$. 
Finally, if we marginalize 
over the previous host halo mass and w$=-1$ ($\Lambda$ cosmology),
we find
$\Omega_m=0.24\pm 0.06$ and $\sigma_{8}=0.83^{+0.11}_{-0.16}$. 

\section*{Acknowledgments} 
We thank Dr. J. Ebrero for providing us with an electronic
version of their clustering results and their XMM survey area-curve and
Dr. N.P. Ross for providing us the electronic version of the SDSS 'QSO 
uniform sample' bias results.
M.P. acknowledges financial support under Mexican government CONACyT grant 2005-49878.


\begin{table}
\caption[]{The best fit cosmological parameters from the likelihood analysis.
}
\tablenotetext{a}{Errors of the fitted parameters 
represent $1\sigma$ uncertainties.}
\tabcolsep 4pt
\begin{tabular}{ccccc} 
\hline
$H_{0}/{\rm kms}^{-1}{\rm Mpc}^{-1}$ & $\Omega_m$& ${\rm w}$&
$\sigma_{8}$& $M/10^{13}h^{-1}M_{\odot}$\\ \hline 
71 &  $0.27^{+0.03}_{-0.05}$  & $-0.90^{+0.10}_{-0.16}$ & $0.74^{+0.14}_{-0.12}$ &  $2.50^{+0.50}_{-1.50}$ \\
74 &  $0.26^{+0.04}_{-0.05}$  & $-0.92^{+0.08}_{-0.14}$ & $0.72^{+0.16}_{-0.14}$&  $2.50^{+0.50}_{-1.50}$ \\ 
71 &  $0.24\pm 0.06$  & $-1$ & $0.83^{+0.11}_{-0.16}$&  $2.50$ \\ \hline


\end{tabular}
\end{table}

\end{document}